\documentclass[11pt]{article}
\usepackage[latin1]{inputenc}
\usepackage[english]{babel}
\usepackage[namelimits]{amsmath}
\usepackage{amssymb}
\usepackage{amsmath}
\usepackage{amsthm}

\begin{document}

\title{Generating Anisotropic Fluids From Vacuum Ernst Equations}

\author{Stefano Viaggiu
\\
Dipartimento di Matematica Universit\'a di Roma ``Tor Vergata'',\\
Via della Ricerca Scientifica, 1, I-00133 Roma, Italy\\
E-mail: viaggiu@mat.uniroma2.it}
\date{\today}\maketitle
\begin{abstract}
Starting with any stationary axisymmetric vacuum metric,
we build anisotropic fluids. With the help of
the Ernst method, the basic equations are derived together with the expression
for the energy-momentum tensor and with the equation of state compatible 
with the field equations. The method is presented by using different 
coordinate systems: the cylindrical 
coordinates $\rho, z$ and the oblate 
spheroidal ones.
A class of 
interior solutions 
matching with stationary axisymmetric 
asymptotically flat vacuum solutions 
is found in
oblate spheroidal coordinates.
The solutions presented satisfy the three energy conditions.   
\end{abstract}	
Keywords : Anisotropic pressure; Ernst equations; Interior solutions.\\
PACS Numbers: 04.20.-q, 04.20.Jb\\
\section*{Introduction}
In literature, many solutions generating technique exist.
Ehlers \cite{1} first showed how it is possible to construct new stationary
exterior solutions and interior ones starting from static vacuum solutions and
applying certain conformal transformations to auxiliary metrics defined
on three-dimensional manifolds. These are one-parameter family of
solutions. Geroch \cite{2} showed that one can obtain an
infinite-parameter family. Further, Xanthopoulos \cite{3} gave a technique
for generating, from any one-parameter family of vacuum solutions, a
two-parameter family. Physically, it is very important to find interior 
solutions describing isolated rotating fluids. Methods have been developed
\cite{4,10}
to build a physically admissible source for a rotating body, but a complete
physically reasonable
space-time for an isolated body is still lacking. Particularly interesting 
is the technique described in \cite{11} to generate perfect fluid solutions
by using a method similar to the vacuum generating method present in
\cite{12,13}. The equations of state compatible with this method   
are $\epsilon=p\;,\;\epsilon+3p=0$. This techinque has been also applied in  
\cite{14} and generalized  in \cite{15} to anisotropic fluids.
Anisotropic fluids are having an increasing interest since they are considered
physically reasonable and appropriate in systems with higher density
and therefore for very compact objects as the core of  neutron stars.
Anisotropic fluids have been studied, for example, in \cite{16,let}.\\
In this paper we present a simple technique to obtain anisotropic  
fluids from any vacuum solution of Ernst \cite{19} equations. A 
technique similar to the one outlined 
in this paper has been used \cite{bel} in a
cosmological context, with two spacelike Killing vectors, to generate 
inhomogeneous cosmological solutions.
In particular, we consider solutions representing isolated bodies together with
a study of the matching conditions
with any stationary asymptotically flat vacuum
metric. All the solutions presented satisfy the three energy conditions
\cite{H,tr}.\\
In sections 1 and 2 we write down the basic equations. In section 3
our method is presented and 
the energy-momentum tensor compatible with the field
equations is studied together with a 
discussion of the energy conditions.
In section 4 we analyze a class of solutions available with our method.\\
Section 5 collects some final remarks.

\section{Basic Equations}
Our starting point is the line element for a stationary 
axisymmetric space-time:
\begin{equation}
ds^2= e^{v}\left[{(dx^1)}^2+{(dx^2)}^2\right]+ld{\phi}^2+2md\phi dt
-fdt^2 ,
\label{1}
\end{equation}
with:
\begin{equation}
v=v(x^1, x^2)\;,\;l=l(x^1, x^2)\;,\;m=m(x^1, x^2)\;,\;f=f(x^1, x^2),
\label{2}
\end{equation}
where $x^1, x^2$ are spatial coordinates, $x^3=\phi$ is an angular coordinate
and $x^4=t$ is a time coordinate. Further, we have \cite{20,21}:
\begin{equation}
fl+m^2={\rho}^2,
\label{3}
\end{equation}
where $\rho$ is the radius in a cylindrical coordinate system. 
With the only non-vanishing components of the 
energy-momentum tensor $T_{\mu\nu}$ given by $T_{11}, T_{22}, 
T_{33}, T_{34}, T_{44}$, the field equations 
$R_{\mu\nu}=-(T_{\mu\nu}-\frac{g_{\mu\nu}}{2}T)$, where
 $T=g^{\mu\nu}T_{\mu\nu}$, are:
\begin{eqnarray}
& &R_{11}=\frac{1}{2}(v_{11}+v_{22})+\frac{{\rho}_{11}}{\rho}-
\frac{v_1{\rho}_1}{2\rho}+\frac{v_2{\rho}_2}{2\rho}-
\frac{1}{2{\rho}^2}(f_1 l_1+{m}_{1}^{2})=\nonumber\\
& &=\frac{e^v}{2}T-T_{11},\label{6}\\
& &R_{22}=\frac{1}{2}(v_{11}+v_{22})+\frac{{\rho}_{22}}{\rho}+
\frac{v_1{\rho}_1}{2\rho}-\frac{v_2{\rho}_2}{2\rho}-
\frac{1}{2{\rho}^2}(f_2 l_2+{m}_{2}^{2})=\nonumber\\
& &=\frac{e^v}{2}T-T_{22},\label{7}\\
& &R_{12}=\frac{{\rho}_{12}}{\rho}-\frac{v_2{\rho}_1}{2\rho}-
\frac{v_1{\rho}_2}{2\rho}-\frac{1}{4{\rho}^2}{\Pi}^{\prime}=0,\label{8}\\
& &R_{33}=\frac{e^{-v}}{2}\left[{\tilde{\nabla}}^2 l+\frac{l}{{\rho}^2}
(f_1 l_1+f_2 l_2+{m}_{1}^{2}+{m}_{2}^{2})\right]=\frac{l}{2}T-T_{33},
\label{9}\\
& &R_{34}=\frac{e^{-v}}{2}\left[{\tilde{\nabla}}^2 m+
\frac{m}{{\rho}^2}(f_1 l_1+f_2 l_2+{m}_{1}^{2}+{m}_{2}^{2})\right]=
\frac{m}{2}T-T_{34},\label{10}\\
& &R_{44}=\frac{e^{-v}}{2}\left[-{\tilde{\nabla}}^2 f-
\frac{f}{{\rho}^2}(f_1 l_1+f_2 l_2+{m}_{1}^{2}+{m}_{2}^{2})\right]=\nonumber\\
& &= -\frac{f}{2}T-T_{44},\label{11}
\end{eqnarray}
where:
\begin{eqnarray}
& &{\Pi}^{\prime}=f_1 l_2+f_2 l_1+2 m_1 m_2,\nonumber\\
& &{\tilde{\nabla}}^2={\partial}_{\alpha\alpha}^2-
\frac{{\rho}_{\alpha}}{\rho}{\partial}_{\alpha}. \label{a1}
\end{eqnarray}
A summation over $\alpha$ is implicit in (\ref{a1}) with $\alpha =1,2,$
i.e. $x^1, x^2$ and subindices denote partial derivatives.
From (\ref{6}) and (\ref{7}) we obtain:
\begin{equation}
-\frac{{\rho}_{11}}{\rho}+\frac{{\rho}_{22}}{\rho}+
\frac{v_1{\rho}_1}{\rho}-\frac{v_2{\rho}_2}{\rho}+
\frac{1}{2{\rho}^2}{\Sigma}^{\prime}=T_{11}-T_{22},
\label{12}
\end{equation}
where:
\begin{equation}
{\Sigma}^{\prime}=f_1 l_1-f_2 l_2+{m}_{1}^{2}-{m}_{2}^{2}
\label{afr}
\end{equation}
From the equations (\ref{8}) and (\ref{12}) we can obtain a first order
differential system for $v$. In the vacuum  
(\ref{6}), (\ref{7}), (\ref{8}) reduce to two
independent equations. In fact, 
when (\ref{3}), (\ref{6}), (\ref{12}) and (\ref{9})-(\ref{11}) are 
used, the equation $R_{11}+R_{22}$ becomes an identity.
Therefore, the relevant equations for the vacuum are (\ref{8}) and
(\ref{12}) together with (\ref{9})-(\ref{11}) (with $T_{\mu\nu}=0$).\\   
In the vacuum, the equations 
(\ref{9})-(\ref{11}) permit to know $f, l, m$
and, as a result, the equations
(\ref{8}) and (\ref{12}) can be completely solved.
Conversely, when matter is present, equations (\ref{6})-(\ref{8})
do not reduce to two independent equations . 
Therefore, an equation can be obtained by 
adding (\ref{6}) with (\ref{7}). 
Further, the vacuum equations, i.e. (\ref{6})-(\ref{11}) with 
$T_{\mu\nu}=0$, imply that: 
\begin{equation}
{\rho}_{\alpha\alpha}=0.
\label{H}
\end{equation} 
Condition (\ref{H}) is retained also in the presence of matter since it is an
assumption greatly simplifying the calculations.\\ 
Finally, we get:
\begin{eqnarray}
& &v_1=c-\frac{[{\rho}_1{\Sigma}^{\prime}+{\rho}_2{\Pi}^{\prime}]}
 {2\rho({\rho}_{1}^{2}+{\rho}_{2}^{2})}+\frac{\rho{\rho}_1(T_{11}-T_{22})}
  {({\rho}_{1}^{2}+{\rho}_{2}^{2})},\label{13}\\
& &v_2=d+\frac{[{\rho}_2{\Sigma}^{\prime}-{\Pi}^{\prime}{\rho}_1]}
 {2\rho({\rho}_{1}^{2}+{\rho}_{2}^{2})}-
\frac{\rho{\rho}_2(T_{11}-T_{22})}
      {({\rho}_{1}^{2}+{\rho}_{2}^{2})},\label{14}\\
& &v_{11}+v_{22}-\frac{1}{2{\rho}^2}(f_{\alpha}l_{\alpha}+{m}_{\alpha}^2)=
Te^{v}-T_{11}-T_{22},\label{a2}\\ 
& &c=\frac{[2{\rho}_{12}{\rho}_2+{\rho}_1({\rho}_{11}-{\rho}_{22})]}
 {({\rho}_{1}^{2}+{\rho}_{2}^{2})}\;,\;
   d=\frac{[2{\rho}_{12}{\rho}_1-{\rho}_2({\rho}_{11}-{\rho}_{22})]}
 {({\rho}_{1}^{2}+{\rho}_{2}^{2})},\label{16}\\
& &e^{-v}[{\tilde{\nabla}}^2 f+\frac{f}{{\rho}^2}(f_{\alpha}l_{\alpha}+
m_{\alpha}^2)]=fT+2T_{44},\label{19}\\
& &e^{-v}[{\tilde{\nabla}}^2 l+\frac{l}{{\rho}^2}(f_{\alpha}l_{\alpha}+
m_{\alpha}^2)]=lT-2T_{33},\label{20}\\
& &e^{-v}[{\tilde{\nabla}}^2 m+\frac{m}{{\rho}^2}(f_{\alpha}l_{\alpha}+
m_{\alpha}^2)]=mT-2T_{34}.\label{21}
\end{eqnarray}

\section{Ernst-like form for the field equations}
In this section
we write down the relevant field equations in a form similar 
to the one of vacuum Ernst equations.\\  
First of all, thanks to (\ref{3}), we can eliminate $l$ from the field
equations and so the equation (\ref{20}),
with the help of (\ref{19}) and
(\ref{21}), becomes an identity. Further, we 
introduce the functions $\gamma, \omega$ with $e^{2\gamma}=fe^v,
m=f\omega$. After made these simplifications, we can introduce the Ernst 
potential $\Phi$, where:
\begin{equation}
{\Phi}_1=\frac{f^2}{\rho}{\omega}_2\;\;\;,\;\;\;
{\Phi}_2=-\frac{f^2}{\rho}{\omega}_1.
\label{27}
\end{equation} 
When (\ref{27}) is used, the equation (\ref{21}) is an identity. To obtain
another field equation, we impose the integrability condition
for (\ref{27}), i.e. ${\omega}_{12}={\omega}_{21}$. 
Further, to take advantage of the Ernst method \cite{19} for the vacuum,
the simplest assumption is:
\begin{eqnarray}
& &2T_{44}+fT=0,\label{34}\\
& &\omega fT-2T_{34}=0,\label{35}\\
& &\frac{({\rho}^2-{\omega}^2 f^2)}{f}T-2T_{33}=0,\label{36}.
\end{eqnarray}
The conditions (\ref{34})-(\ref{36}) permit us to set to zero the right hand side
of the equations (\ref{19})-(\ref{21}). In practice, the equations involving 
the functions $f, \Phi$ are the same of the vacuum case.
As a result, the field
equations (\ref{13})-(\ref{14}) and  (\ref{19})-(\ref{21}) become
\begin{eqnarray}
& &f{\nabla}^2 f +{\Phi}_{\alpha}^{2}-{f}_{\alpha}^{2}=0,\label{28}\\
& &f{\nabla}^2\Phi-2{f}_{\alpha}{\Phi}_{\alpha}=0,\label{29}\\
& &{\gamma}_1=-\frac{{\rho}_1{\Sigma}+{\rho}_2\Pi}
{4\rho({\rho}_{1}^{2}+{\rho}_{2}^{2})}+\frac{c}{2}+
\frac{\rho{\rho}_1(T_{11}-T_{22})}
{2({\rho}_{1}^{2}+{\rho}_{2}^2)},\label{30}\\
& &{\gamma}_{2}=\frac{{\rho}_2\Sigma-\Pi{\rho}_1}
{4\rho({\rho}_{1}^{2}+{\rho}_{2}^2)}+\frac{d}{2}-
\frac{\rho{\rho}_2(T_{11}-T_{22})}
{2({\rho}_{1}^{2}+{\rho}_{2}^{2})},\label{31}\\
& &\Sigma=\frac{{\rho}^2}{f^2}(f_{2}^{2}-f_{1}^{2})+
f^2[{\omega}_{1}^{2}-
    {\omega}_{2}^{2}]\label{32}\\
& &\Pi=-2\frac{{\rho}^2}{f^2}f_1 f_2+2f^2{\omega}_1{\omega}_2,\label{33}\
\end{eqnarray}
where:
\begin{equation}
{\nabla}^2={\partial}_{\alpha\alpha}^2+
\frac{{\rho}_{\alpha}}{\rho}{\partial}_{\alpha},
\label{mar}
\end{equation}
with the line element:
\begin{equation}
ds^2=f^{-1}\left[e^{2\gamma}\left({(dx^1)}^2+{(dx^2)}^2\right)+{\rho}^2
{d\phi}^2\right]-f{(dt-\omega d\phi)}^2.
\label{37}
\end{equation} 
The line element (\ref{37}) is written in the so called Papapetrou
gauge \cite{20}.\\
Finally, with the help of (\ref{13}) and (\ref{14}), the
equation (\ref{a2}) looks as follows:
\begin{equation}
\frac{e^{2\gamma}}{f}T-T_{11}-T_{22}=\frac{({\rho}_{1}^{2}-{\rho}_{2}^{2})(T_{11}-T_{22})
+\rho{\rho}_{1}{(T_{11}-T_{22})}_{1}-\rho{\rho}_{2}
{(T_{11}-T_{22})}_{2}}{({\rho}_{1}^{2}+{\rho}_{2}^{2})}.
\label{25}
\end{equation}
In the next section we present our method.

\section{The method}
\subsection{Integrability and energy conditions}
Thanks to equations (\ref{34})-(\ref{36}), the equations 
(\ref{28})-(\ref{29}) have the same structure of the vacuum Ernst equations
with the Ernst potential $\Phi$ given by equations (\ref{27}).
Obviously, since the equations (\ref{30})-(\ref{31}) contain terms
involving the energy-momentum tensor, to integrate our system of equations
we need a further integrability condition.\\ 
We get this integrability condition
by setting  ${\gamma}_{12}={\gamma}_{21}$. 
We read:
\begin{equation}
2{\rho}_2{\rho}_{1}(T_{11}-T_{22})+\rho{\rho}_{1}{(T_{11}-T_{22})}_{2}+
\rho{\rho}_{2}{(T_{11}-T_{22})}_{1}=0.
\label{26}
\end{equation}
To obtain the equation (\ref{26}), we have used the equations
(\ref{28})-(\ref{29}).  
In this way, we can obtain
an interior solution with the same two-metric, spanned by the Killing
vectors ${\partial}_{t}$ and ${\partial}_{\phi}$, 
of the vacuum seed metric. 
In practice, we obtain anisotropic fluids by ``gauging'' the metric
function $\gamma$ of the seed vacuum solution. 
Obviously, is always possible to
build anisotropic space-times by varying the metric coefficients. But, in this
manner, we obtain an energy-momentum tensor with a non-appealing 
expression. In fact, since $g_{12}=0$,
if $R_{12}\neq 0$ then $T_{\mu\nu}$ has 
the component $T_{12}\neq 0$, and this does not allow to write a simple
expression for $T_{\mu\nu}$.
Further, if $f, \omega$ are not
solutions of the vacuum Ernst equations,
then equation (\ref{8}) is not easy to solve.\\ 
The next step is to study the form of the energy-momentum tensor
allowed by the field equations.
To this purpose, we assume that the four-velocity $V^{\mu}$
of the fluid is $V^{\mu}={\delta}^{\mu}_{4}/\sqrt{f}$, i.e.
we assume that our coordinates are co-rotating with the fluid
(see \cite{Herre}).
With the line element (\ref{37}), we can write $T_{\mu\nu}$ in the 
form:
\begin{equation}
T_{\mu\nu}=(\epsilon+p_1)V_{\mu}V_{\nu}+
p_1g_{\mu\nu}+(p_3-p_1)K_{\mu}K_{\nu}+
(p_2-p_1)S_{\mu}S_{\nu}
\label{b1}
\end{equation}  
where $\epsilon$ is the mass-energy density, $p_1, p_2, p_3$ are the
principal stresses, $V_{\mu}, S_{\mu}, K_{\mu}$ are four-vectors
satisfying:
\begin{equation}
V^{\mu}V_{\mu}=-1,\;K^{\mu}K_{\mu}=S^{\mu}S_{\mu}=1,\;
V^{\mu}K_{\mu}=V^{\mu}S_{\mu}=K^{\mu}S_{\mu}=0,
\end{equation}
being, with respect to (\ref{37}):
\begin{eqnarray}
K_{\mu}&=&\left[0\;,\;0\;,\;\frac{\rho}{\sqrt{f}}\;,\;0\right],
\nonumber\\
S_{\mu}&=&\left[0\;,\;\frac{e^{\gamma}}{\sqrt{f}}\;,\;0\;,\;0\right],
\nonumber\\
V_{\mu}&=&\left[0\;,\;0\;,\;
\omega\sqrt{f}\;,\;
-\sqrt{f}\right].
\label{b2}
\end{eqnarray}
With respect to the tensor (\ref{b1}), the eigenvalues ${\lambda}$
(see \cite{tr}) are given by the roots of the equation:  
\begin{equation}
|T_{\mu\nu}-\lambda g_{\mu\nu}|=0,
\label{sec}
\end{equation}
i.e.
$(p_1-\lambda)(p_2-\lambda)(\epsilon+\lambda)(p_3-\lambda)=0$.
Therefore, we obtain:
${\lambda}_{1}=p_1, {\lambda}_{2}=p_2, {\lambda}_3=p_3, 
{\lambda}_4=-\epsilon $.\\ Furthermore, we must impose the energy
conditions \cite{H,tr}, that in our notations are, for the weak
energy condition ($i=1,2,3$):
\begin{equation}
-{\lambda}_4\geq 0\;\;,\;\;-{\lambda}_4+{\lambda}_i\geq 0;
\label{b3}
\end{equation}
for the strong energy condition:
\begin{equation}
-{\lambda}_4+\sum {\lambda}_i\geq 0\;\;,\;\;-{\lambda}_4+{\lambda}_i\geq 0;
\label{b4}
\end{equation}
and for the dominant energy condition:
\begin{equation}
-{\lambda}_4\geq 0\;\;,\;\;{\lambda}_4\leq {\lambda}_i\leq-{\lambda}_4 .
\label{b5}
\end{equation}
We can write the equations (\ref{34})-(\ref{36}) in terms of the
principal stresses and of the mass-energy density. We obtain:
\begin{eqnarray}
& &(\epsilon +p_3)({\rho}^2+f^2{\omega}^2)-(p_1+p_2)({\rho}^2-f^2{\omega}^2)
= 0 ,\label{b6}\\
& &\omega [\epsilon +p_1+p_2+p_3] = 0,\label{b7}\\
& &\epsilon +p_1+p_2+p_3 = 0 .\label{b8}
\end{eqnarray}
Equations (\ref{b7}) and (\ref{b8}) are equivalent. 
When (\ref{b8}) is put in (\ref{b6}), we have
$p_1=-p_2,\;\epsilon =-p_3$. 

\subsection{Starting steps}

Let us summarize all the relevant equations we need:
\begin{eqnarray}
& &p_1=-p_2=p\;\;\;,\;\;\;\epsilon = -p_3 ,\label{b9}\\
& &T_{11}-T_{22}=\frac{2p}{f}e^{2\gamma} ,\label{b10}\\
& &\left(2{\rho}_2{\rho}_1+{\rho}{\rho}_1{\partial}_{2}+
{\rho}{\rho}_{2}{\partial}_{1}\right)\left[T_{11}-T_{22}\right]=0 ,
\label{b11}\\
& &\frac{\left({\rho}_{1}^{2}-{\rho}_{2}^{2}+{\rho}{\rho}_{1}{\partial}_{1}-
{\rho}{\rho}_{2}{\partial}_{2}\right)\left[T_{11}-T_{22}\right]}
{({\rho}_{1}^{2}+{\rho}_{2}^2)} = \frac{e^{2\gamma}}{f}(p_3-\epsilon) .
\label{b12}
\end{eqnarray}
together with the equations (\ref{30}) and (\ref{31}). 
A simple step by step
procedure to solve the system (\ref{b9})-(\ref{b12}) is
the following. 
\begin{enumerate}
\item First of all, we choose a solution of the vacuum Ernst equations
by means of the functions $(f, \omega)$. 
\item We specify the initial coordinates by
the relation with $\rho, z$.
\item We integrate the equation (\ref{b11}) to find
$T_{11}-T_{22}$ as a function of the choosen coordinates.
\item We calculate, with the obtained function
$T_{11}-T_{22}$ and
by means of equation (\ref{b10}),
the function $p$ as a function of the  
unknown function $\gamma$ and of the
known function $f$.
\item We put the solution obtained for $T_{11}-T_{22}$ 
in (\ref{b12})
to calculate, by means of the second equation of (\ref{b9}),
the function $\epsilon$,
and therefore all the principal stresses together
with the mass-energy density $\epsilon$ are found in terms of
the unknown function $\gamma$ and of the
known function $f$. 
\item As a final step, we can substitute 
the expression for $T_{11}-T_{22}$ in (\ref{30}) and (\ref{31}) and calculate
the metric function $\gamma$.
The condition (\ref{b11}) guaranties the
integrability of (\ref{30}) and (\ref{31}).
\item As a result, starting with 
choosen functions $f,\omega$ as known solutions of the vacuum Ernst equations,
we generate the solution with $\gamma, f, \omega$ with the line element
given by (\ref{37}) and with a non-vanishing energy-momentum tensor
given by (\ref{b1}).
\end{enumerate}
Note that, since the equation
(\ref{b11}) does not depend on $\gamma$, our method is self-consistent. 
Hence, the equations 
for $\gamma$ can be solved without ambiguity.\\ 
Our method cannot describe perfect fluid solutions.
In fact, for a perfect fluid solution $p_1=p_2=p_3$ and therefore
$T_{11}-T_{22}=0$. In this case, the field equations imply that 
$p_1=p_2=p_3=\epsilon=0$. Therefore the only perfect fluid solution
allowed with our method is the vacuum one.\\
Further, note that no restrictions 
are made on $f$, and $\omega$: they are only 
solutions of the vacuum Ernst equations.
Furthermore, in the static limit $\omega=0$, we obtain interior solutions 
with equation of state (\ref{b9}), (\ref{b10}), (\ref{b12}).\\ 
In the next section we present our method with some physically interesting
examples.
\section{Application of the method}
\subsection{First example: Cylindrical coordinates}
Starting with the line element (\ref{37}) with $x^1=\rho,
x^2=z$, we have ${\rho}_{1}=1, {\rho}_{2}=0$. 
In the chosen coordinates,
the most general solution for (\ref{b11}) is:
\begin{equation}
T_{11}-T_{22}=F(\rho),\label{c3}
\end{equation}
where $F(\rho)$ is an arbitrary regular function.\\
Expressions (\ref{b10}) and (\ref{b12}) become:
\begin{eqnarray}
& &\epsilon = -\frac{f}{2e^{2\gamma}}(F+\rho F_{\rho}),\label{Ag1}\\
& &p = \frac{f}{2e^{2\gamma}}F \label{Ag2}.
\end{eqnarray} 
It is a simple matter to verify that the only regular class of functions
for $F(\rho)$ satisfying all the energy conditions are the
non-positive ones that are monotonically decreasing. As an example, we assume
$F(\rho)=c(-\rho-{\rho}^2)$, and as a result the generated solution is:
\begin{eqnarray}
& &\epsilon =
\frac{cf}{2e^{2\gamma}}[2\rho+3{\rho}^2]\nonumber\\
& &2\gamma=2{\gamma}_0+c{\rho}^3
\left[-\frac{\rho}{4}-\frac{1}{3}\right]+c\alpha, \label{g1}
\end{eqnarray} 
with $c,\alpha$ constant and $c>0$.\\
If we search for interior solutions to match with exterior vacuum ones, then 
we need solutions with a static surface and with vanishing 
hydrostatic pressure. To this purpose, the continuity of the first and
the second fundamental form \cite{9,10} on a surface with
$\rho=R=const.$ with vanishing pressure requires that:
\begin{equation}
{\gamma}_0(R)=\gamma(R)\;\;,\;\;
{\gamma}_{0\rho}(R)={\gamma}_{\rho}(R).
\label{R54}
\end{equation}
It is easy to see that, in order to satisfy the energy conditions and (\ref{R54}),
$F(\rho)$ must be positive and singular on the axis at $\rho=0$.
To see this, we start from expressions (\ref{Ag1}), (\ref{Ag2}).
First, suppose that $F(0)<0$. Then, the energy conditions are satisfied
only if
\begin{equation}
-F-\rho{F}_{\rho}\geq -F . 
\label{c7}
\end{equation}
Expression (\ref{c7}) implies that $F_{\rho}\leq 0$, and thus $F$ is decreasing
in a neighbourhood of $\rho =0$. But, in this way, $F$ cannot be zero at some
radius $\rho =R$.\\
Conversely, suppose that $F(0)\geq 0$. To satisfy the energy conditions, we
must have:
\begin{equation} 
-\rho F_{\rho}\geq 2F .
\label{c8}
\end{equation}
Thus, from (\ref{c8}) we deduce that
$F$ is decreasing in a neighbourhood of $\rho=0$.
But, thanks to (\ref{c8}),
if $F$ is regular for $0\leq\rho\leq R$, then
we conclude that $F(0)=0$ and therefore $F$ cannot be zero at some radius
$R$.\\
Concluding, the only way to satisfy both energy conditions 
and (\ref{R54}) is to choose
$F$ to be positive and irregular on the axis. Equation (\ref{c8}) implies that,
in a neighbourhood of $\rho =0$, $F$ must show the following behaviour: 
$F(\rho)\geq\frac{1}{{\rho}^2}$.\\
Perhaps the most simple class of solutions that we can consider is:
\begin{equation}
F(\rho)=\frac{c}{{\rho}^{k^2+1}}{(R-\rho)}^{s^2+1},
\label{d1}
\end{equation}
where $c>0$ and $|k|\geq 1, |s|\geq 1$. It is easy to see that the solution
(\ref{d1}) satisfies all the energy conditions and (\ref{R54}). 
After integrating the 
equations for $\gamma$
we get:
\begin{equation}
2\gamma =2{\gamma}_0+c\int \frac{{(R-\rho)}^{s^2+1}}{{\rho}^{k^2}}d\rho
+z(k,s),
\label{dd2}
\end{equation}
where $z(k,s)$ is an integration constant chosen to satisfy
the first of
equations (\ref{R54}). Generally, the integral  (\ref{dd2}) involves 
expressions in terms of hypergeometric functions.\\
Expression (\ref{dd2}) for $\gamma$ can be potentially singular on the
axis. However, we have to choose a seed vacuum metric. Thus, we could choose
an expression for ${\gamma}_0$
such that, in a neighbourhood of $\rho =0$, the expression (\ref{dd2}) be regular.
This can be accomplished by taking a vacuum seed solution with an 
appropriate singular expression
for ${\gamma}_0$ on the axis.
We do not enter into such a discussion, but
only mention the fact that the Lewis
\cite{21} class of solutions are not appropriate.\\
Concerning the solution (\ref{d1}), the most simple example we can consider is:
\begin{equation}
F(\rho)=\frac{c}{{\rho}^2}{(R-\rho)}^2.
\label{m3}
\end{equation}
With (\ref{m3}), after integrating the field equations, our interior 
solutions, matching smoothly on $\rho=R$ with any
vacuum solution and  satisfying all the energy conditions are:
\begin{eqnarray}
& &p_{\rho}=-p_{z}=p\;\;,\;\;\epsilon=-p_{\phi},\nonumber\\
& &2\gamma=2{\gamma}_{0}+c\left[\frac{1}{2}{\rho}^2-2\rho R+R^2\ln(\rho)+
\frac{3}{2}R^2-R^2\ln(R)\right],\nonumber\\
& &p=\frac{cf}{2{\rho}^{(2+cR^2)}e^{(2{\gamma}_0)}}
{(R-\rho)}^2e^{-c[\frac{1}{2}{\rho}^2-2\rho R+\frac{3}{2}R^2-R^2\ln(R)]},
\nonumber\\
& &\epsilon=
     \frac{cf}{2{\rho}^{(2+cR^2)}e^{(2{\gamma}_0)}}(R^2-{\rho}^2)
            e^{-c[\frac{1}{2}{\rho}^2-2\rho R+\frac{3}{2}R^2-R^2\ln(R)]}.
\label{c12}
\end{eqnarray}
As an example, if we choose for
${\gamma}_0$ an expression that, for $\rho\rightarrow 0$,
looks as follows:
\begin{equation}
e^{(2{\gamma}_0)}\simeq\frac{H(z)}{{\rho}^{(cR^2)}}+higher\;\;orders,
\label{g5}
\end{equation}
where $H(z)$  is a regular positive non-vanishing
function and $f\rightarrow {\rho}^2$ when  
$\rho\rightarrow 0$,
then all the expressions in (\ref{c12}) for
$\gamma$, $p$ and $\epsilon$
are  regular on the axis.\\      
To conclude this subsection, we write down a class of interior solutions
matching with  vacuum ones in such a way that the  matter is in the region
$R\leq\rho<\infty$, 
while the region $0\leq\rho<R$ is filled with  vacuum.
In this case it is a simple matter to see that, to satisfy the energy conditions,
$F(\rho)$ must be a regular negative 
decreasing function when $\rho\geq R$.
As a simple class of solutions, we have:
\begin{eqnarray}
& &F=-c{(\rho-R)}^{\alpha}\;\;,\;\;\alpha\geq 1\;,\;c>0\nonumber\\
& &\epsilon=\frac{cf}{2e^{2\gamma}}{(\rho-R)}^{\alpha-1}
(\rho-R+\rho\alpha)\nonumber\\
& &2\gamma = 2{\gamma}_0-c\int\rho{(\rho-R)}^{\alpha}d\rho.
\label{se2}
\end{eqnarray}
When $\alpha$ is an integer, the solution (\ref{se2}) has a simple
expression.

\subsection{Second example : Oblate spheroidal coordinates} 
We consider oblate spheroidal coordinates defined in terms of
the cylindrical ones $\rho$, $z$ as: 
$\rho=\cosh\mu\cos\theta\;,\;z=\sinh\mu\sin\theta$, where 
$0\leq\mu<\infty $ and
$-\frac{\pi}{2}\leq\theta\leq\frac{\pi}{2}$.\\
Looking for metrics describing isolated objects with a vanishing hydrostatic
pressure surface, we have the following solution:
\begin{equation}
T_{11}-T_{22}=\frac{c(s^2{\cosh}^2\mu-k{\cos}^2\theta-ks^2{\cos}^2\theta)}
{{\cosh}^4\mu},
\label{d2}
\end{equation}
where $k, s, c$ are arbitrary constants ($s>0$).\\
Thus, we get:
\begin{eqnarray}
& &p=\frac{f}{2e^{2\gamma}}\frac{1}{{\cosh}^4\mu}(T_{11}-T_{22}),\;
p_{\mu}=-p_{\theta}=p,\; \epsilon=-p_{\phi},\label{d3}\\
& &\epsilon =\frac{fc}{2e^{2\gamma}}\frac{1}{{\cosh}^4\mu}
\left[s^2{\cosh}^2\mu-3k(1+s^2){\cos}^2\theta\right].\label{d4}
\end{eqnarray}
The energy conditions can be fulfilled if $c>0, k<0$.
But, in this way, $p$ cannot  vanish on some boundary surface.
Remember that it is not necessary for $p_{\phi}$ to vanish at some 
surface to identify the boundary of the source region.
To this purpose, we must take in (\ref{d3}) and (\ref{d4}) 
$c<0, k>0$. Hence, the energy conditions are satisfied in the region where
\begin{equation}
\cosh\mu\leq\frac{\cos\theta}{s}\sqrt{2k(1+s^2)}.
\label{d5}
\end{equation}
The surface of vanishing hydrostatic pressure is:
\begin{equation}
\cosh\mu=\frac{\cos\theta}{s}\sqrt{k(1+s^2)}=
A\cos\theta.
\label{d6}
\end{equation}
Expression (\ref{d6}) represents effectively the equation of a 
boundary surface
only if:
\begin{equation}
\frac{s^2}{k(1+s^2)}<1,
\label{d8}
\end{equation}
or $A>1$.
In fact, by expressing the equation (\ref{d6}) in terms of 
cylindrical coordinates, we read
\begin{equation}
z^2=\frac{1}{A}(A\rho-1)(A-\rho).
\label{gen}
\end{equation}
The surface (\ref{gen}) is a toroid (genus=1)
with inner and outer radii $r_1,r_2$ given by
$r_1=1/A, r_1 r_2=1$, with $\rho\in [r_1,r_2]$
and $z\in [(1-A^2)/(2A),(A^2-1)/(2A)]$. 
When $A=1$ the surface degenerates to a circle of radius $\rho=1$
with $z=0$.\\
As a consequence, thanks to (\ref{d5}) and (\ref{d8}), in the region enclosed
by the surface (\ref{d6}), the energy conditions follow. 
Integrating the equations for $\gamma$, we get:  
\begin{eqnarray}
& &\gamma =\frac{c}{2}\left[(s^2-k(1+s^2))
\ln(\frac{\sqrt{{\cosh}^2\mu-{\cos}^2\theta}}{\cosh\mu})
-\frac{1}{2}k(1+s^2)\frac{{\cos}^2\theta}{{\cosh}^2\mu}\right]+\nonumber\\
& &+{\gamma}_0+\frac{c}{2}\alpha,\label{d7}
\end{eqnarray}
with $\alpha$ a constant. Expression (\ref{d7}) has a ring singularity
at $\mu=0, \theta=0$, i.e. $\rho=1, z=0$. Thanks to (\ref{d8}), this 
singularity lies in the interior of the surface (\ref{d6})
($\cosh\mu < \frac{\cos\theta}{s}\sqrt{k(1+s^2)}$). Consequently, the metric
can be potentially singular at that place.
However, to
understand the nature of this singularity, a seed vacuum metric must be
specified. The situation is similar to the one for cylindrical
coordinate. 
Also in this case, we must find an expression
for ${\gamma}_0$ such that the expression (\ref{d7}) is regular, but
we do not enter into the search.\\
It is a simple matter to verify that by choosing
\begin{equation}
\alpha = \frac{s^2}{2}+\frac{1}{2}(ks^2+k-s^2)\ln\frac{(k+ks^2-s^2)}
{k(1+s^2)},
\label{d9}
\end{equation} 
we have, on the boundary $S$ of (\ref{d6}):
\begin{equation}
{\gamma}_0(S)=\gamma(S)\;,\;{\gamma}_{0 i}(S)={\gamma}_{i}(S)\;,\;
i=\mu,\theta.
\label{d10}
\end{equation}
Therefore, our interior metric is $C^1$ on the boundary surface $S$
and thus
it can be matched smoothly to any stationary axisymmetric
asymptotically flat solution with
$f,\omega, {\gamma}_0$. Note that, thanks to (\ref{d8}), expression
(\ref{d9}) is real.\\ Further, the principal stress $p_{\mu}$ is always
positive.\\ 

\subsection{Third example: The Kerr metric}
We choose now, as seed metric, the Kerr one.
After writing the solution (\ref{d7}) in the Boyer-Lindquist coordinates 
(see \cite{BL,v}), we get:
\begin{eqnarray}
& &ds^2=\Sigma\left(d{\theta}^2+\frac{dr^2}{\Delta}\right)e^{F}+
(r^2+a^2){\sin}^2\theta d{\phi}^2-dt^2+\nonumber\\
& &+\frac{2mr}{\Sigma}{\left(dt+a{\sin}^2\theta d\phi\right)}^2 , \label{w1}\\
& &\Sigma =r^2+a^2{\cos}^2\theta\;,\;
\Delta =r^2+a^2-2mr ,\nonumber\\
& & F=c\left[(s^2-k(1+s^2))
\ln\frac{\sqrt{\Delta-{\sin}^2\theta}}{\sqrt{\Delta}}-
\frac{1}{2}k(1+s^2)\frac{{\sin}^2\theta}{\Delta}+\alpha\right],\nonumber
\end{eqnarray}
with $\alpha$ given by (\ref{d9}). The interior metric written in
the Boyer-Lindquist coordinates can be extended to all the values  
of the parameters $a^2, m^2$ allowed by the Kerr solution.
In particular, when $a^2\leq m^2$, the surface of zero pressure 
$\Delta =\frac{k(1+s^2)}{s^2}{\sin}^2\theta$ 
generally does not describe a toroidal
surface, but a closed surface passing trought the $z$ axis. Further, solution
(\ref{w1}) can be defined when $\frac{\Delta-{\sin}^2\theta}{\Delta}<0$
by setting, for example,
$\frac{c[s^2-k(1+s^2)]}{2}=2n$, where 
$n$ is a positive integer.\\
When $a^2>m^2$, the surface 
$\Delta= \frac{k(1+s^2)}{s^2}{\sin}^2\theta$ becomes
a toroidal rotational surface as (\ref{d6}) (see \cite{v}).\\
It is interesting to note that,
with the coordinates used in (\ref{w1}), when $a^2>1+m^2$ ($\Delta>1$),
the ring singularity of (\ref{d7}) disappears.
In this case, the only singularity for the global spacetime
(both interior and exterior metric) is the ring of the vacuum Kerr solution.
For
\begin{equation}
m^2+1<a^2<\frac{k(1+s^2)}{s^2}
\label{erc}
\end{equation} 
the Kerr ring lies in the matter region.
Otherwise, the Kerr ring belongs to the vacuum exterior Kerr solution.\\
Finally, note that $\Sigma e^{F}=\frac{e^{2\gamma}}{f}>0$, in such a way that
energy conditions follow within the surface (\ref{d6})
($c<0, k>0$).
\section{Conclusions}
In this paper we have presented a simple technique to obtain
anisotropic fluids starting from any vacuum solution of the Ernst
equations. The equation of state compatible with our method is:
$p_1=-p_2,\;\epsilon=-p_3$.\\
In \cite{15}, anisotropic solutions with 
$\epsilon+p_1+p_2+p_3=0$ have been obtained with the help of Geroch  
\cite{13} transformations applied to matter space-times. In this way,
starting with matter space-times endowed with 
almost a Killing vector and with the appropriate equation of state,
we can obtain new solutions adding twist to
the seed space-time. For the solutions so obtained, the equation of state
is the one of the seed metric
with the matter parameters scaled by a common factor.\\
The equations of state compatible with the method of Krisch and Glass are:
$3\epsilon+p_2=0\;,p_1=p_3=-\epsilon$ for a 
space-time admitting  a space-like Killing vector,
and $\epsilon=p_2\;,\;p_2=-p_1=-p_3$ for a 
space-time with a time-like Killing vector.
However, to apply our method, 
no seed matter space-time is needed, 
but only vacuum stationary
axially symmetric solutions. Moreover, note that our equation of state
$p_1=-p_2, \epsilon=-p_3$ contains the one of Krisch and Glass 
given by $\epsilon=p_2, p_2=-p_1=-p_3$ as a limiting case.\\
It is worth noticing that with our method we can always obtain
isolated sources matching with asymptotically flat solutions.\\
We have analyzed the problem of joining the generating solutions with 
exterior vacuum ones. By using
cylindrical coordinates, we are able to
match our anisotropic metrics, on the boundary surfaces $\rho=R$,
with all vacuum solutions
in such a way that all the energy conditions are satisfied. 
Further, an interior solution  
is obtained representing
an isolated body with an ``unusual'' but
physically acceptable equation of state matching with
any stationary axisymmetric asymptotically flat solution.\\
In both cases, the regularity of the solutions so obtained are discussed.
Finally, it can be noticed that the solution (\ref{w1}) could be used 
to describe extreme astrophysical situations where a Kerr black hole
is surronded by non usual matter.

\end{document}